\documentclass[letter]{article}
\usepackage{graphicx}
\usepackage{amsmath}

\begin{document}

%\begin{titlepage}

\centerline{\bf CONFINEMENT OF THE FIELD ANGULAR MOMENTUM}

\bigskip

\centerline{\bf Vladimir Dzhunushaliev}

\medskip

\centerline{\small Department of Phys. and Microel. Engineering, }

\centerline{\small Kyrgyz-Russian Slavic University, Kievskaya
Str. 44,}

\centerline{\small Bishkek,  720021, Kyrgyz Republic}

\medskip

\centerline{\small\tt dzhun@hotmail.kg}

\bigskip

\centerline{Received March 20, 2005}

\medskip

\begin{abstract}
It is shown that the quantum SU(3) gauge theory can be
approximately reduced to U(1) gauge theory with broken gauge
symmetry and interacting with scalar fields. The scalar fields are
some approximations for 2 and 4-points Green's functions of $A^{1,
\ldots , 7}_\mu$ gauge potential components. The remaining gauge
potential component $A^8_\mu$ is the potential for U(1) Abelian
gauge theory. It is shown that reduced field equations have a
regular solution. The solution presents a quantum bag in which
$A^8_\mu$ color electromagnetic field is confined. This field
produces a field angular momentum which can be equal to $\hbar$.
It is supposed that the obtained solution can be considered as a
model of glueball with spin $\hbar$. In this model the glueball
has an asymmetry. The same asymmetry may have the nucleon which
can be measured experimentally.\\

\end{abstract}

%\end{titlepage}

\section{Introduction}

One of the problems of the nucleon spin structure is the origin of
the orbital angular momentum of the gluon field. In this paper, we
offer the following model of the orbital angular momentum in a
quantum bag. The SU(3) gauge potential $A^B_\mu \in SU(3)$ ($B=1,
\ldots , 8, \mu = 0,1,2,3$) can be decomposed on $A^a_\mu \in
SU(2) \subset SU(3)$ ($a=1,2,3$), $A^8_\mu \in U(1) \subset SU(3)$
and $A^m_\mu \in SU(3)/(SU(2) \times U(1)$ ($m=4,5,6,7$). The
non-perturbative interaction between quantum components $A^a_\mu$
and $A^m_\mu$ leads to the appearance of a pure quantum bag
\cite{dzhun1}. We will show that in this bag one can place color
electric and magnetic fields (arising from $A^8_\mu$) in such a
way that an orbital angular momentum appears which is confined in
the bag.
\par
The idea presented here that the gluonic field $A^B_\mu$ with
different $B$ may have a different behaviour is not a new idea.
For example, in Ref. \cite{Simonov:2005ir} the similar idea is
stated on the language of ``valence gluon field $a_\mu$ and
background field $B_\mu$'' and as a result the behaviour of field
correlators $D$ and $D_1$ is obtained at small and large distances
for perturbative and non-perturbative parts.
\par
Recently \cite{Gripaios}-\cite{Dudal} it is shown that the
condensate $\left\langle A^B_\mu A^{B\mu} \right\rangle$ may play
very important role in QCD. In the present paper we will show that
this condensate is absolutely necessary to describe the quantum
bag, in which some almost-classical color field is confined.

\section{$SU(3) \rightarrow SU(2) + U(1) + Coset$ decomposition}

In this section the decomposition of $SU(3)$ gauge field to the
subgroup $SU(2) \times U(1)$ is defined. Starting with the $SU(3)$
gauge group with generators $T^B$, we define the $SU(3)$ gauge
fields $\mathcal{A}_\mu=\mathcal{A}^B_\mu T^B$. Let $U(1) \times
SU(2)$ be a subgroup of SU(3) and $SU(3)/(U(1) \times SU(2))$ is a
coset. Then the gauge field $\mathcal{A}_\mu$ can be decomposed as
follows:
\begin{eqnarray}
  \mathcal{A}_\mu & = & \mathcal{A}^B_\mu T^B = A^a_\mu T^a +
  b_\mu T^8 + A^m_\mu T^m ,
\label{sec2-10a}\\
  A^a_\mu & \in & SU(2), \quad A^8_\mu = b_\mu \in U(1)
  \text{ and }  A^m_\mu \in SU(3)/(U(1) \times SU(2))
\label{sec2-10b}
\end{eqnarray}
where the indices $B = 1, \ldots ,8$ are SU(3) indices; $a,b,c
\ldots = 1,2,3$ belongs to the subgroup SU(2) and $m,n, \ldots
=4,5,6,7$ belongs to the coset $SU(3)/(U(1) \times SU(2))$. On
this basis, the field strength can be decomposed as
\begin{equation}
  \mathcal{F}^B_{\mu\nu} T^B = \mathcal{F}^a_{\mu\nu}T^a + \mathcal{F}^8_{\mu\nu}T^8 +
  \mathcal{F}^m_{\mu\nu}T^m
\label{sec2-20}
\end{equation}
where
\begin{eqnarray}
  \mathcal{F}^a_{\mu\nu}  = F^a_{\mu\nu} + g f^{amn}A^m_\mu A^n_\nu
   \; \in SU(2) ,
\label{sec2-30a}\\
  F^a_{\mu\nu}  =  \partial_\mu A^a_\nu - \partial_\nu A^a_\mu +
  g \epsilon^{abc}A^b_\mu A^c_\nu \; \in SU(2) ,
\label{sec2-30b}\\
  \mathcal{F}^8_{\mu\nu}  =  h_{\mu \nu} +
  g f^{8mn} A^m_\mu A^n_\nu  \; \in U(1) ,
\label{sec2-30c}\\
    h_{\mu \nu}  =  \partial_\mu b_\nu - \partial_\nu b_\mu \; \in U(1) ,
\label{sec2-30f}\\ \nonumber
  \mathcal{F}^m_{\mu\nu}  =  F^m_{\mu\nu} +
  g f^{mna} \left( A^a_\nu A^n_\mu - A^a_\mu A^n_\nu \right) \\+
  g f^{mn8} \left( b_\nu A^n_\mu - b_\mu A^n_\nu \right)
  \;  \in SU(3)/(U(1) \times SU(2)) ,
\label{sec2-30d}\\ \nonumber
  F^m_{\mu\nu}  =  \partial_\mu A^m_\nu - \partial_\nu A^m_\mu +
  g f^{mnp} A^n_\mu A^p_\nu  \; \\
   \in SU(3)/(U(1) \times SU(2)) ,
\label{sec2-30e}
\end{eqnarray}
where $f^{ABC}$ are structure constants of SU(3), $\epsilon^{abc}
= f^{abc}$ are structure constants of SU(2) and $g$ is the
coupling constant.
\par
For the non-perturbative quantization we will apply a modification
of the Heisenberg quantization technique to the SU(3) Yang-Mills
equations. In quantizing these classical system via Heisenberg's
method \cite{heis} one first replaces classical fields by field
operators $\mathcal{A}^B _{\mu} \rightarrow
\hat{\mathcal{A}}^B_\mu$. This yields non-linear, coupled,
differential equations for the field operators. One then uses
these equations to determine expectation values for the field
operators $\hat{\mathcal{A}}^B_\mu$ ({\it e.g.} $\langle
\hat{\mathcal{A}}^B_\mu \rangle$, where $\langle \cdots \rangle =
\langle Q | \cdots | Q \rangle$ and $| Q \rangle$ is some quantum
state). One can also use these equations to determine the
expectation values of operators that are built up from the
fundamental operators $\hat{\mathcal{A}}^B_\mu$. For example, the
``electric'' field operator $\mathcal{\hat E}^B_z = \partial _0
\hat{\mathcal{A}}^B_z -
\partial_z \hat{\mathcal{A}}^B_0 +
g f^{BCD} \mathcal{\hat A}^C_0 \mathcal{\hat A}^D_z$ giving the
expectation $\langle \mathcal{\hat E}^a_z \rangle$. The simple
gauge field expectation values, $\langle \mathcal{\hat A}_\mu (x)
\rangle$, are obtained by taking the expectation of the operator
version of Yang-Mills equations with respect to some quantum state
$| Q \rangle$. One problem in using these equations to obtain
expectation values like $\langle \hat{\mathcal{A}}^B_\mu \rangle$,
is that these equations involve not only powers or derivatives of
$\langle \hat{\mathcal{A}}^B_\mu \rangle$ ({\it i.e.} terms like
$\partial_\alpha \langle \hat{\mathcal{A}}^B_\mu \rangle$ or
$\partial_\alpha \partial_\beta \langle \hat{\mathcal{A}}^B_\mu
\rangle$) but also contain terms like $\mathcal{G}^{mn}_{\mu\nu} =
\langle \mathcal{\hat A}^B_\mu \mathcal{\hat A}^C_\mu \rangle$.
Starting with the operator version of Yang-Mills equations one can
generate an operator differential equation for the product
$\hat{\mathcal{A}}^B_\mu \hat{\mathcal{A}}^C_\nu$ thus allowing
the determination of the Green's function
$\mathcal{G}^{mn}_{\mu\nu}$. However this equation will in turn
contain other, higher-order Green's functions. Repeating these
steps leads to an infinite set of equations connecting Green's
functions of ever increasing order. This procedure is very similar
to the field correlators approach in QCD (for a review, see
\cite{giacomo}). In Ref. \cite{simonov} a set of self coupled
equations for such field correlators is given. This construction,
leading to an infinite set of coupled, differential equations,
does not have an exact analytical solution and so must be handled
using some approximation.

\section{Derivation of an effective Lagrangian}

\subsection{Basic assumptions for the reduction}

It is evident that a full and exact quantization is impossible in
this case. Thus we have to look for some simplification in order
to obtain equations which can be analyzed. Our basic aim is to cut
off the infinite equations set using some simplifying assumptions.
Our quantization procedure will derive from the Heisenberg method
in which we will take the expectation of the Lagrangian rather
than for the equations of motions. Thus we will obtain an
effective Lagrangian rather than approximate equations of motion.
For this purpose we have to have ansatz for the following 2 and
4-points Green's functions:
 $$\left\langle A^a_\mu(x) A^b_\nu(y) \right\rangle,\quad
 \left\langle A^m_\mu(x)A^n_\nu(y) \right\rangle,$$
 $$\left\langle A^a_\alpha (x)A^b_\beta (y) A^m_\mu(z) A^n_\nu(u)
 \right\rangle,\quad
 \left\langle A^a_\alpha (x) A^b_\beta (y) A^c_\mu(z) A^d_\nu(u)
\right\rangle,$$
 $$\left\langle A^m_\alpha (x) A^n_\beta (y)
A^p_\mu(z) A^q_\nu(u) \right\rangle, \quad \left\langle b_\alpha
(x) b_\beta (y) A^m_\mu(z) A^n_\nu(u) \right\rangle.$$
 The field $b_\mu$ remains to be almost classical field. Now we would
like to list the assumptions necessary for the simplification of 2
and 4-points Green's functions.
\begin{enumerate}
\item
\label{ass1}
      The gauge field components $A^8_\mu = b_\mu$
      belonging to the small subgroup $U(1)$ are in an ordered phase. Mathematically
      this means that
\begin{equation}
  \left\langle b_\mu (x) \right\rangle  =(b_{\mu} (x))_{cl}.
\label{sec3-10}
\end{equation}
      The subscript means that this is a classical field. Thus we are
      treating these components as effectively classical gauge fields
      in the first approximation. In Ref. \cite{Li:2004zu}, similar idea on
      the decomposition of initial degrees of freedom to almost-classical and
      quantum degrees of freedom is applied to provide calculation of the
      $\left\langle  A^B_\mu A^{B\mu} \right\rangle$
      condensate. There the condensate is a constant but in fact
      in our paper we propose the method which allow us to calculate the condensate
      varying in the space.
\item
\label{ass2}
            The gauge field components
            $A^a_\mu \in SU(2)$ (a=1,2,3) belonging to the subgroup $SU(2)$,
            $A^m_\mu \in SU(3)/(U(1) \times SU(2))$ (m=4,5, ... , 7) belonging to the coset
      $SU(3)/(U(1) \times SU(2)$ are in
      a disordered phase (in other words, a condensate), but have
      non-zero energy. In mathematical terms this means that
            \begin{eqnarray}
                \left\langle A^{a_1}_{\mu_1}(x_1) \cdots A^{a_{2n+1}}_{\mu_{2n+1}} (x_{2n+1})
                \right\rangle = 0,
               \text{ but }
              \left\langle A^{a_1}_{\mu_1}(x_1) \cdots A^{a_{2n}}_{\mu_{2n}} (x_{2n})
              \right\rangle \neq 0
            \label{sec3-20}\\
                \left\langle A^{m_1}_{\mu_1}(x_1) \cdots A^{m_{2n+1}}_{\mu_{2n+1}} (x_{2n+1})
                \right\rangle = 0,
               \text{ but }
              \left\langle A^{m_1}_{\mu_1}(x_1) \cdots A^{m_{2n}}_{\mu_{2n}} (x_{2n})
              \right\rangle \neq 0 .
            \label{sec3-30}
            \end{eqnarray}
      We suppose that
\begin{enumerate}
    \item
    \label{ass2a}
    \begin{equation}
        \left\langle A^a_\mu (x) A^b_\nu (y) \right\rangle =
        - \eta_{\mu \nu} f^{apm} f^{bpn} \phi^m(x) \phi^n(y)
        \label{sec3-40};
        \end{equation}
    \item
    \label{ass2b}
    $$
    \left\langle A^m_\mu (x) A^n_\nu (y) \right\rangle =
          - \eta_{\mu \nu} \left[
            f^{mpa} f^{npb} \phi^a(x) \phi^b(y) +
            \delta^{mn} \psi(x) \psi(y)\right.
    $$
        \begin{equation}
           \left. + \alpha f^{amn} \psi(x) \psi(y)
          \right]
        \label{sec3-50}
        \end{equation}
        where $\alpha$ is some for the time undefined constant.
\end{enumerate}
\item
\label{ass3}
            There is the correlation between quantum phases $A^a_\mu$ and $A^m_\mu$
$$
\left\langle
                \left(
                    A^{a_1}_{\mu_1}\left( x_1) \right) \ldots A^{a_n}_{\nu_n}   \left( x_n \right)
                \right)
              \left(
                A^m_\alpha (y) \ldots A^m_\beta (z)
              \right)
              \right\rangle
$$
            \begin{equation}
               =
              k_n
              \left\langle
                \left(
                    A^{a_1}_{\mu_1}\left( x_1) \right) \ldots A^{a_n}_{\nu_n}   \left( x_n \right)
                \right)
                \right\rangle
                \left\langle
              \left(
                A^m_\alpha (y) \ldots A^m_\beta (z)
              \right)
              \right\rangle
            \label{sec3-60}
            \end{equation}
            where the coefficient $k_n$ describes the correlation between these phases and
            depends on the number of operators.
\item
\label{ass4}
            There is the correlation between ordered (classical) and disordered (quantum) phases
$$
\left\langle
                \left(
                    b_{\mu_1} \ldots b_{\mu_n}
                \right)
              \left(
                A^a_\alpha \ldots A^b_\beta
              \right)
              \left(
                A^m_\gamma \ldots A^n_\delta
              \right)
              \right\rangle
$$
                \begin{equation}
                    =
              r_n \left( b_{\mu_1} \ldots b_{\mu_n} \right)
                \left\langle
              \left(
                A^a_\alpha \ldots A^b_\beta
              \right)
              \left(
                A^m_\gamma \ldots A^n_\delta
              \right)
              \right\rangle
            \label{sec3-70}
            \end{equation}
            where the coefficients $r_n$ describe the correlation between these phases.
\item
\label{ass5}
            The 4-point Green's function can be expressed via 2-points Green's
            functions
            \begin{enumerate}
                \item
                $$
\left\langle
                A^m_\mu(x) A^n_\nu(y) A^p_\alpha(z) A^q_\beta(u)
            \right\rangle
                $$
                \label{ass5a}
                \begin{equation}
                            \begin{split}
             = &
            \lambda_1
            \biggl(
                \left\langle A^m_\mu A^n_\nu \right\rangle
                \left\langle A^p_\alpha A^q_\beta \right\rangle -
                \mu_1^2 \eta_{\alpha\beta} \delta^{pq} \left\langle A^m_\mu A^n_\nu \right\rangle -
                \mu_1^2 \eta_{\mu\nu} \delta^{mn} \left\langle A^p_\alpha A^q_\beta \right\rangle + \\
                &\left\langle A^m_\mu A^p_\alpha \right\rangle
                \left\langle A^n_\nu A^q_\beta \right\rangle -
                \mu_1^2 \eta_{\nu\beta} \delta^{nq} \left\langle A^m_\mu A^p_\alpha \right\rangle -
                \mu_1^2 \eta_{\mu\alpha} \delta^{mp} \left\langle A^n_\nu A^q_\beta \right\rangle + \\
                &\left\langle A^m_\mu A^q_\beta \right\rangle
                \left\langle A^n_\nu A^p_\alpha \right\rangle -
                \mu_1^2 \eta_{\nu\alpha} \delta^{np} \left\langle A^m_\mu A^q_\beta \right\rangle -
                \mu_1^2 \eta_{\mu\beta} \delta^{mq} \left\langle A^n_\nu A^p_\alpha \right\rangle
            \biggl);
            \end{split}
            \label{sec3-82}
            \end{equation}
            \item
            \label{ass5b}
$$\left\langle
                A^a_\mu(x) A^b_\nu(y) A^c_\alpha(z) A^d_\beta(u)
            \right\rangle
$$
            \begin{equation}
            \begin{split}
             = &
            \lambda_2
            \biggl(
                \left\langle A^a_\mu A^b_\nu \right\rangle
                \left\langle A^c_\alpha A^d_\beta \right\rangle -
                \mu_1^2 \eta_{\alpha\beta} \delta^{cd} \left\langle A^a_\mu A^b_\nu \right\rangle -
                \mu_1^2 \eta_{\mu\nu} \delta^{ab} \left\langle A^c_\alpha A^d_\beta \right\rangle + \\
                &\left\langle A^a_\mu A^c_\alpha \right\rangle
                \left\langle A^b_\nu A^d_\beta \right\rangle -
                \mu_1^2 \eta_{\nu\beta} \delta^{bd} \left\langle A^a_\mu A^c_\alpha \right\rangle -
                \mu_1^2 \eta_{\mu\alpha} \delta^{ac} \left\langle A^b_\nu A^d_\beta \right\rangle + \\
                &\left\langle A^a_\mu A^d_\beta \right\rangle
                \left\langle A^b_\nu A^c_\alpha \right\rangle -
                \mu_1^2 \eta_{\nu\alpha} \delta^{ad} \left\langle A^b_\mu A^c_\beta \right\rangle -
                \mu_1^2 \eta_{\mu\beta} \delta^{bc} \left\langle A^a_\nu A^d_\alpha \right\rangle
            \biggl);
            \end{split}
            \label{sec3-84}
            \end{equation}
            \item
            \label{ass5c}
            \begin{equation}
                \begin{split}
                \left\langle
                    b_\mu(x) b_\nu(y) A^m_\alpha(z) A^n_\beta(u)
                \right\rangle = &
                r_2 b_\mu b_\nu
                \left\langle
                    A^m_\alpha A^n_\beta
                \right\rangle -
                b_\mu b_\nu \delta^{mn} M_{\alpha \beta}
            \end{split}
            \label{sec3-84a}
            \end{equation}
            where $M_{\alpha \beta}$ is some constant matrix.
            \end{enumerate}
\end{enumerate}
It is necessary to note that: (a) according to the assumptions \eqref{ass2} and \eqref{ass5}
\textit{the scalar fields $\phi^{a,m}$ are not the classical fields} but describe
2 and 4-points Green's function of the gauge potential $A^{m,a}_\mu$; (b)
we consider the static case only, i.e. all Green's functions do not depend on the
time; (c) the assumption \eqref{ass5} means that schematically
$\left\langle A^4 \right\rangle = \left\langle A^2 \right\rangle
\left\langle A^2 \right\rangle - \mu^2 \left\langle A^2 \right\rangle + M$
and that the initial system loses some symmetry (gauge symmetry in our case).

\subsection{The first step. $SU(3) \rightarrow SU(2) + U(1) + coset$ reduction}

Our main aim is to show that the quantum SU(3) gauge theory in
some physical situations can be approximately reduced to U(1) +
scalar fields theory. In this section we will show that $SU(3)
\rightarrow U(1) + coset$ reduction can be made by two steps. On
the first step we will decompose $SU(3) \rightarrow SU(2) + U(1) +
coset$ and on the second step $SU(2) + U(1) \rightarrow U(1) +
coset$. Thus our aim is to calculate
\begin{equation}
    \left\langle
        \mathcal{F}^B_{\mu\nu} \mathcal{F}^{B\mu\nu}
    \right\rangle =
    \left\langle
        \mathcal{F}^a_{\mu\nu} \mathcal{F}^{a\mu\nu}
    \right\rangle +
    \left\langle
        \mathcal{F}^8_{\mu\nu} \mathcal{F}^{8\mu\nu}
    \right\rangle +
    \left\langle
        \mathcal{F}^m_{\mu\nu} \mathcal{F}^{m\mu\nu}
    \right\rangle
\label{sec3-85}
\end{equation}

\subsubsection{Calculation of $\left\langle \mathcal{F}^a_{\mu\nu} \mathcal{F}^{a\mu\nu} \right\rangle $}

We begin by calculating
\begin{equation}
    \left\langle
        \mathcal{F}^a_{\mu\nu} \mathcal{F}^{a\mu\nu}
    \right\rangle =
    \left\langle
        F^a_{\mu\nu} F^{a\mu\nu}
    \right\rangle +
    2 g f^{amn}
    \left\langle
        F^a_{\mu\nu} A^{m\mu} A^{n\nu}
    \right\rangle +
    g^2 f^{amn} f^{apq}
    \left\langle
        A^m_\mu A^n_\nu A^{p\mu} A^{q\nu}
    \right\rangle .
\label{sec3-90}
\end{equation}
According to the assumption \eqref{ass2a},
\begin{equation}
    \left\langle
        F^a_{\mu \nu} A^{m \mu} A^{n\nu}
    \right\rangle = 0
\label{sec3-100}
\end{equation}
since $\left\langle F^a_{\mu \nu} \right\rangle$ is antisymmetric
tensor while $A^{m \mu} A^{n\nu}$ is a symmetric one. According to
the assumption \eqref{ass5a} one can calculate (the details can be
found in Ref. \cite{dzhun1})
\begin{equation}
    \left\langle
        A^m_\mu(x) A^n_\nu(x) A^p_\alpha(x) A^q_\beta(x)
    \right\rangle =
        \lambda_1 g^2 \left[ \frac{9}{4} \left( \phi^a \phi^a \right)^2 -
        18 \mu_1^2 \phi^a \phi^a
    \right] .
\label{sec3-120}
\end{equation}
Finally,
\begin{equation}
    \left\langle
        \mathcal{F}^a_{\mu \nu} \mathcal{F}^{a\mu \nu}
    \right\rangle =
    \left\langle
        F^a_{\mu \nu} F^{a\mu \nu}
    \right\rangle + \lambda_1 g^2 \left[ \frac{9}{4} \left( \phi^a \phi^a \right)^2 -
        18 \mu_1^2 \phi^a \phi^a
    \right]
\label{sec3-130}
\end{equation}

\subsubsection{Calculation of $\left\langle \mathcal{F}^m_{\mu\nu} \mathcal{F}^{m\mu\nu} \right\rangle $}

The calculation of this term we begin by calculating
$$
\left\langle
        \mathcal{F}^m_{\mu\nu} \mathcal{F}^{m\mu\nu}
    \right\rangle
$$
\begin{equation}\
\begin{split}
     = &
    \left\langle
        \left( \partial_\mu A^m_\nu - \partial_\nu A^m_\mu \right)^2
    \right\rangle +
    g^2 f^{mna} f^{mpb}
    \left\langle
        \left( A^n_\mu A^a_\nu - A^n_\nu A^a_\mu \right)
        \left( A^{p\mu} A^{b\nu} - A^{p\nu} A^{b\mu} \right)
    \right\rangle + \\
    &
    g^2 f^{mn8} f^{mp8}
    \left\langle
        \left( A^n_\mu b_\nu - A^n_\nu b_\mu \right)
        \left( A^{p\mu} b^\nu - A^{p\nu} b^\mu \right)
    \right\rangle + \\
    &
    2g f^{mna}
    \left\langle
        \left( \partial_\mu A^m_\nu - \partial_\nu A^m_\mu \right)
    \right\rangle
    \left\langle
        \left( A^{n\mu} A^{a\nu} - A^{n\nu} A^{a\mu} \right)
    \right\rangle + \\
    &
    2g f^{mn8}
    \left\langle
        \left( \partial_\mu A^m_\nu - \partial_\nu A^m_\mu \right)
    \right\rangle
    \left\langle
        \left( A^{n\mu} b^\nu - A^{n\nu} b^\mu \right)
    \right\rangle + \\
    &
    2g^2 f^{mna} f^{mp8}
    \left\langle
        \left( A^n_\mu A^a_\nu - A^n_\nu A^a_\mu \right)
    \right\rangle
    \left\langle
        \left( A^{p\mu} b^\nu - A^{p\nu} b^\mu \right)
    \right\rangle
\end{split}
\label{sec3-140}
\end{equation}
The calculations using the assumptions \eqref{ass2a},
\eqref{ass3}, \eqref{ass4} and \eqref{ass5c} give us
\begin{eqnarray}
    \left\langle
        \left( \partial_\mu A^m_\nu - \partial_\nu A^m_\mu \right)
        \left( \partial^\mu A^{m\nu} - \partial^\nu A^{m\mu} \right)
    \right\rangle =
    - 6 \left[
        \left( \partial_\mu \phi^a \right)^2 + 4
        \left( \partial_\mu \psi \right)^2
    \right] ,
\label{sec3-150}\\
        2g f^{mna}
    \left\langle
        \left( \partial_\mu A^m_\nu - \partial_\nu A^m_\mu \right)
        \left( A^{n\mu} A^{a\nu} - A^{n\nu} A^{a\mu} \right)
    \right\rangle =
    -6 g k_1 \epsilon^{abc} \left( \partial_\mu \phi^a \right) A^{b\mu} \phi^c ,
\label{sec3-160}\\
        2g f^{mn8}
    \left\langle
        \left( \partial_\mu A^m_\nu - \partial_\nu A^m_\mu \right)
        \left( A^{n\mu} b^\nu - A^{n\nu} b^\mu \right)
    \right\rangle =
    18 g\alpha r_1 b^\mu \left( \partial_\mu \psi \right) \psi ,
\label{sec3-170}\\ \nonumber
    g^2 f^{mna} f^{mpb} \left\langle
        \left( A^n_\mu A^a_\nu - A^n_\nu A^a_\mu \right)
        \left( A^{p\mu} A^{b\nu} - A^{p\nu} A^{b\mu} \right)
    \right\rangle \\=
    -6 g^2 k_2 \left( A^a_\mu A^{a\mu} \right)
    \left[
        \frac{1}{4} \left( \phi^b \phi^b \right) + \psi^2
    \right] ,
\label{sec3-180}\\ \nonumber
        g^2 f^{mn8} f^{mp8}
    \left\langle
        \left( A^n_\mu b_\nu - A^n_\nu b_\mu \right)
        \left( A^{p\mu} b^\nu - A^{p\nu} b^\mu \right)
    \right\rangle\\ =
    -6 g^2 r_2 b_\mu b^\mu
    \left(
        \frac{3}{4} \phi^a \phi^a + 3 \psi^2
    \right) +
       6g^2 b_\mu b_\nu \left( -\eta^{\mu \nu} M^\alpha_\alpha +M^{\mu \nu} \right)    ,
\label{sec3-190}\\
        2g^2 f^{mna} f^{mp8}
    \left\langle
        \left( A^n_\mu A^a_\nu - A^n_\nu A^a_\mu \right)
        \left( A^{p\mu} b^\nu - A^{p\nu} b^\mu \right)
    \right\rangle   = 0 ,
\label{sec3-200}\\
    g^2 f^{amn} f^{apq}
    \left\langle
        A^m_\mu A^n_\nu A^{p\mu} A^{q\nu}
    \right\rangle =
    \lambda_1 g^2 \left[
        \frac{9}{4} \left( \phi^a \phi^a \right)^2 - 18 \mu_1^2 \phi^a \phi^a
    \right],
\label{sec3-210}\\
    g^2 f^{8mn} f^{8pq}
    \left\langle
        A^m_\mu A^n_\nu A^{p\mu} A^{q\nu}
    \right\rangle =
    \lambda_1 g^2 \left[
        \frac{9}{4} \left( \phi^a \phi^a \right)^2 - 18 \mu_1^2 \phi^a \phi^a
    \right] .
\label{sec3-220}
\end{eqnarray}
Finally,
\begin{equation}
\begin{split}
    \left\langle
        \mathcal{F}^m_{\mu \nu} \mathcal{F}^{m\mu \nu}
    \right\rangle =
    - 6 \left[
        \left( \partial_\mu \phi^a \right)^2 + 4
        \left( \partial_\mu \psi \right)^2
    \right]\\
    -6 g^2 k_2 \left( A^a_\mu A^{a\mu} \right)
    \left[
        \frac{1}{4} \left( \phi^b \phi^b \right) + \psi^2
    \right] -
        6 g^2 r_2 b_\mu b^\mu
    \left(
        \frac{3}{4} \phi^a \phi^a + 3 \psi^2
    \right)\\
    -6 g k_1 \epsilon^{abc} \left( \partial_\mu \phi^a \right) A^{b\mu} \phi^c +
    18 g\alpha r_1 b^\mu \left( \partial_\mu \psi \right) \psi
\end{split}
\label{sec3-240}
\end{equation}

\subsubsection{Calculation of $\left\langle \mathcal{F}^8_{\mu\nu} \mathcal{F}^{8\mu\nu} \right\rangle $}

Analogously we have
\begin{equation}\
\begin{split}
    \left\langle
        \mathcal{F}^8_{\mu\nu} \mathcal{F}^{8\mu\nu}
    \right\rangle =
    \left( h_{\mu \nu} \right)^2 + 2g h_{\mu \nu} f^{8mn}
    \left\langle A^{m\mu} A^{n\nu} \right\rangle +
    g^2 f^{8mn} f^{8pq}
    \left\langle
        A^m_\mu A^n_\nu A^{p\mu} A^{q\nu}
    \right\rangle  \\
        =\left( h_{\mu \nu} \right)^2 +
    \lambda_1 g^2 \left[
        \frac{9}{4} \left( \phi^a \phi^a \right)^2 - 18 \mu_1^2 \phi^a \phi^a
    \right]
\end{split}
\label{sec3-250}
\end{equation}
as $h_{\mu \nu}$ is the antisymmetric tensor but
$\left\langle A^{m\mu} A^{n\nu} \right\rangle$ is symmetric one.

\subsection{An effective Lagrangian after the first step}

Using the results of the previous sections we have
\begin{equation}
\begin{split}
    \left\langle
        \mathcal{F}^A_{\mu\nu} \mathcal{F}^{A\mu\nu}
    \right\rangle =
    F^a_{\mu\nu} F^{a\mu\nu} +
    h_{\mu \nu} h^{\mu \nu} \\
        -6 \left\{
        \left( \partial_\mu \phi^a \right)^2 +
        k_1 g \epsilon^{abc} \left( \partial_\mu \phi^a \right) A^{b\mu} \phi^c \right.\\
        \left.+ g^2 \frac{k_2}{4} \left[
            \left( A^a_\mu A^{a\mu} \right) \phi^b \phi^b -
            \left( A^a_\mu \phi^a \right) \left( A^b_\mu \phi^b \right)
        \right]
    \right\}  \\
        -6g^2 \frac{k_2}{4} \left( A^a_\mu \phi^a \right) \left( A^b_\mu \phi^b \right)
    + \lambda_1 g^2 \left[
        \frac{9}{2} \left( \phi^a \phi^a \right)^2 - 36 \mu_1^2 \phi^a \phi^a
    \right]  \\
        -24 \left[
        \left( \partial_\mu \psi \right)^2 -
        \frac{3}{4} g \alpha r_1 \left( \partial_\mu \psi \right) b^\mu \psi +
        \frac{3}{4} g^2 r_2 b_\mu b^\mu \psi^2
    \right]\\ -
    6g^2 \left[
        k_2 \left( A^a_\mu A^{a\mu} \right) +
        \frac{3}{4} r_2 \left( b_\mu b^\mu \right) 
    \right] \psi^2 \\
     - \frac{9}{2} g^2 r_2 \phi^a \phi^a b_\mu b^\mu + 6g^2 b_\mu b_\nu
    \left( -\eta ^{\mu \nu} M^\alpha_\alpha + M^{\mu \nu} \right).
\end{split}
\label{sec3-260}
\end{equation}
One can choose the following for the time undefined parameters
\begin{equation}
    r_2 = \frac{4}{3}, \quad r_1 = \sqrt{r_2} = \frac{2}{\sqrt{3}},
    \quad \alpha = -\frac{4}{\sqrt{3}}
\label{sec3-270}
\end{equation}
and redefine
\begin{equation}
    \phi^a \rightarrow \frac{\phi^a}{\sqrt 3}, \quad
    \psi^a \rightarrow \frac{\psi^a}{\sqrt{12}}, \quad
    \lambda_1 \rightarrow 2\lambda_1, \quad
    \mu^2_1 \rightarrow \frac{\mu^2_1}{12}, \quad
    -\eta_{\mu \nu} M^\alpha_\alpha + M_{\mu \nu} \rightarrow
    \frac{\left( m^2 \right)_{\mu \nu}}{3} .
\label{sec3-280}
\end{equation}
After this we will have the following effective Lagrangian:
\begin{equation}
\begin{split}
    -4 \left\langle \mathcal{L}_{SU(3)} \right\rangle = \left\langle
        \mathcal{F}^A_{\mu\nu} \mathcal{F}^{A\mu\nu}
    \right\rangle =
    F^a_{\mu\nu} F^{a\mu\nu} +
    h_{\mu \nu} h^{\mu \nu} - \\
         2\left\{
        \left( \partial_\mu \phi^a \right)^2 +
        k_1 g \epsilon^{abc} \left( \partial_\mu \phi^a \right) A^{b\mu} \phi^c +
        g^2 \frac{k_2}{4} \epsilon^{abc} \epsilon^{ade} A^b_\mu \phi^c A^{d\mu} \phi^e
    \right\} - \\
        2\left( D_\mu \psi \right)^2 -
    2g^2 \left( A^a_\mu \phi^a \right) \left( A^{b\mu} \phi^b \right) +
    \lambda_1 g^2 \left(
        \phi^a \phi^a - \mu_1^2
    \right)^2  - \lambda_1 g^2 \mu_1^4 - \\
        2g^2 \left[
        A^a_\mu A^{a\mu} + b_\mu b^\mu 
    \right] \psi^2 +
    2g^2 b_\mu b_\nu \left( m^2 \right)^{\mu \nu}
\end{split}
\label{sec3-290}
\end{equation}
where $F^a_{\mu \nu} = \partial_\mu A^a_\nu - \partial_\nu A^a_\mu +
g \epsilon^{abc} A^b_\mu A^c_\nu$ is the field tensor of the nonabelian SU(2) gauge group;
$h_{\mu \nu} = \partial_\mu b_\nu - \partial_\nu b_\mu$ is the tensor of the abelian
U(1) gauge group; $D_\mu \psi = \partial_\mu \psi + g b_\mu \psi$ is the gauge
derivative of a scalar field $\psi$ with respect to the U(1) gauge field $b_\mu$.
\par
It is interesting to note that if we choose
\begin{equation}
    k_1 = 2, \quad k_2 = 4
\label{sec3-300}
\end{equation}
then we will have the $SU(2) + U(1)$ Yang-Mills-Higgs theory with
broken gauge symmetry,
\begin{equation}
\begin{split}
    -4 \left\langle  \mathcal{L}_{SU(3)} \right\rangle = &\left\langle
        \mathcal{F}^A_{\mu\nu} \mathcal{F}^{A\mu\nu}
    \right\rangle =
    F^a_{\mu\nu} F^{a\mu\nu} +
    h_{\mu \nu} h^{\mu \nu}  -
    2 \left( D_\mu \phi^a \right)^2 - \\
    &
    2\left( D_\mu \psi \right)^2 -
    2g^2 \left( A^a_\mu \phi^a \right) \left( A^{b\mu} \phi^b \right) +
    \lambda_1 g^2 \left(
        \phi^a \phi^a - \mu_1^2
    \right)^2  - \lambda_1 g^2 \mu_1^4 - \\
    &
    2g^2 \left[
        \left( A^a_\mu A^{a\mu} + b_\mu b^\mu \right) \psi^2 +
        \left( b_\mu b^\mu \right) \left( \phi^a \phi^a \right)
    \right]
\end{split}
\label{sec3-310}
\end{equation}
where $D_\mu \phi = \partial_\mu \phi^a + g \epsilon^{abc} A^b_\mu \phi^c$
is the gauge derivative with respect to the SU(2) gauge field $A^a_\mu$.

\subsection{The second step. $SU(2) + U(1) \rightarrow U(1) + coset$ decomposition}

Now we will quantize $A^a_\mu$ degrees of freedom. First, we will
calculate the term
$$
 \left\langle  F^a_{\mu \nu} F^{a\mu \nu} \right\rangle =
    \left\langle
        \left( \partial_\mu A^a_\nu - \partial_\nu A^a_\mu \right)^2
    \right\rangle +
    2g \epsilon^{abc}
    \left\langle
        \left( \partial_\mu A^a_\nu - \partial_\nu A^a_\mu \right) A^{b\mu} A^{c\nu}
    \right\rangle
$$
\begin{equation}
    +
    g^2 \epsilon^{abc} \epsilon^{ade}
    \left\langle
        A^b_\mu A^c_\nu A^{d\mu} A^{e\nu}
    \right\rangle .
\label{sec4-10}
\end{equation}
The second term in the rhs of Eq. \eqref{sec4-10} is zero as the
consequence of the assumption \eqref{ass2}: $\left\langle A^3
\right\rangle = 0$. Using this result we have
\begin{equation}
    \left\langle  F^a_{\mu \nu} F^{a\mu \nu} \right\rangle =
    - \frac{9}{2} \left( \partial_\mu \phi^m \right)^2 +
    \lambda_2 g^2
    \left[
        \frac{9}{2} \left( \phi^m \phi^m \right)^2 - 36 \mu^2_2 \phi^m \phi^m
    \right] .
\label{sec4-20}
\end{equation}
The next terms are
\begin{eqnarray}
    \left\langle  A^a_\mu A^{a\mu} \right\rangle \psi^2 &=&
    - 3 \left( \phi^m \phi^m \right) \psi^2,
\label{sec4-30}\\
    \left\langle
        \left( D_\mu \phi^a \right)^2
    \right\rangle +
    g^2 \left\langle
        \left( A^a_\mu \phi^a \right) \left( A^{b\mu} \phi^b \right)
    \right\rangle &=&
    \left( \partial_\mu \phi^a \right)^2 -
    \frac{3k_2}{4} \left( \phi^a \phi^a \right) \left( \phi^m \phi^m \right).
\label{sec4-40}
\end{eqnarray}
Collecting all the terms with $A^a_\mu$ we have
\begin{equation}
\begin{split}
    \left\langle  F^a_{\mu \nu} F^{a\mu \nu} \right\rangle -
    2 \left\langle
        \left( D_\mu \phi^a \right)^2
    \right\rangle -
    2 g^2 \left\langle
        \left( A^a_\mu \phi^a \right) \left( A^{b\mu} \phi^b \right)
    \right\rangle -
    2 \left\langle  A^a_\mu A^{a\mu} \right\rangle \psi^2 = \\
        - \frac{9}{2} \left( \partial_\mu \phi^m \right)^2 +
    \lambda_2 g^2 \left[
        \frac{9}{2} \left( \phi^m \phi^m \right)^2 - 36 \mu_2^2 \left( \phi^m \phi^m \right)
    \right] \\-
    2 \left( \partial_\mu \phi^a \right)^2 +
    \frac{3k_2}{2}g^2 \left( \phi^a \phi^a \right)\left( \phi^m \phi^m \right) +
    6g^2 \left( \phi^m \phi^m \right) \psi^2 .
\end{split}
\label{sec4-50}
\end{equation}
After the redefinition $\phi^m \rightarrow \frac{2}{3} \phi^m,
\mu_2^2 \rightarrow \frac{\mu_2^2}{9}, \lambda_2 \rightarrow
\frac{9}{8}\lambda_2$,
\begin{equation}
\begin{split}
    \text{(All terms with }A^a_\mu ) =
    - 2 \left( \partial_\mu \phi^m \right)^2 - 2 \left( \partial_\mu \phi^a \right)^2\\ +
    \lambda_2 g^2 \left[
        \left( \phi^m \phi^m \right) - \mu_2^2
    \right]^2 - \lambda_2 g^2 \mu_2^4 \\+
        \frac{2k_2}{3}g^2 \left( \phi^a \phi^a \right)\left( \phi^m \phi^m \right) +
    \frac{8}{3}g^2 \left( \phi^m \phi^m \right) \psi^2 .
\end{split}
\label{sec4-60}
\end{equation}

\subsection{An effective Lagrangian}

Finally, we have the following effective Lagrangian:
\begin{equation}
\begin{split}
    - \frac{g^2}{4} \left\langle \mathcal{F}^A_{\mu \nu} \mathcal{F}^{A\mu \nu} \right\rangle = &
    - \frac{1}{4} h_{\mu \nu} h^{\mu \nu} +
    \frac{1}{2} \left( \partial_\mu \phi^a \right)
    \left( \partial^\mu \phi^a \right) +
    \frac{1}{2} \left( \partial_\mu \phi^m \right)
    \left( \partial^\mu \phi^m \right) - \\
    &
    \frac{\lambda_1}{4} \left[
        \left( \phi^a \phi^a \right) - \mu_1^2
    \right]^2 -
    \frac{\lambda_2}{4} \left[
        \left( \phi^m \phi^m \right) - \mu_2^2
    \right]^2 - \frac{\lambda_1}{4} \mu_1^4 - \frac{\lambda_2}{4} \mu_2^4 - \\
    &
    \frac{k_2}{6} \left( \phi^a \phi^a \right) \left( \phi^m \phi^m \right) +
    \left( b_\mu b^\mu \right) \phi^a \phi^a -
    \frac{1}{2} \left( m^2 \right)^{\mu \nu} b_\mu b_\nu
\end{split}
\label{sec4-70}
\end{equation}
where we have redefined $b_\mu \rightarrow b_\mu /g, \phi^{a,m}
\rightarrow \phi^{a,m} /g$ and for the simplicity we consider the
case with $\psi = 0$.
\par
The field equations for this theory are
\begin{eqnarray}
    \partial^\mu \partial_\mu \phi^a &=& - \phi^a
    \left[
        \frac{k_2}{3} \phi^m \phi^m +
        \lambda_1 \left( \phi^a \phi^a - \mu_1^2 \right ) - b_\mu b^\mu
    \right] ,
\label{sec4-90}\\
    \partial^\mu \partial_\mu \phi^m &=& - \phi^m
    \left[
        \frac{k_2}{3} \phi^a \phi^a +
        \lambda_2 \left( \phi^m \phi^m - \mu_2^2 \right )
    \right] ,
\label{sec4-100}\\
    \partial_\nu h^{\mu \nu} &=&
    2 b^\mu \left( \phi^a \phi^a \right ) - \left( m^2 \right)^{\mu \nu} b_\nu
\label{sec4-110}
\end{eqnarray}
It is convenient to redefine
$\phi^{a,m} \rightarrow \sqrt{\frac{3}{k_2}}\phi^{a,m},
\mu_{1,2} \rightarrow \sqrt{\frac{3}{k_2}} \mu_{1,2},
\lambda_{1,2} \rightarrow \frac{k_2}{3} \lambda_{1,2}$ and then
\begin{eqnarray}
    \partial^\mu \partial_\mu \phi^a &=& - \phi^a
    \left[
        \phi^m \phi^m +
        \lambda_1 \left( \phi^a \phi^a - \mu_1^2 \right ) - b_\mu b^\mu
    \right] ,
\label{sec4-120}\\
    \partial^\mu \partial_\mu \phi^m &=& - \phi^m
    \left[
        \phi^a \phi^a +
        \lambda_2 \left( \phi^m \phi^m - \mu_2^2 \right )
    \right] ,
\label{sec4-130}\\
    \partial_\nu h^{\mu \nu} &=&
    \frac{6}{k_2} b^\mu \left( \phi^a \phi^a \right ) - \left( m^2 \right)^{\mu \nu} b_\nu
\label{sec4-140}
\end{eqnarray}
Let us note that here we have undefined parameters
$\lambda_{1,2}, \mu_{1,2}, (m^2)_{\mu \nu}, k_2$. In principle these parameters
have to be defined using an \emph{exact} non-perturbative quantization procedure,
for example, path integration.

\section{Numerical solution}

We will search for the solution in the following form:
\begin{eqnarray}
    \phi^a \left( r, \theta \right) &=&
    \frac{\phi\left( r, \theta \right)}{\sqrt{3}} , \quad a=1,2,3 ,
\label{sec5-10}\\
  \phi^m \left( r, \theta \right) &=&
  \frac{\chi \left( r, \theta \right)}{2} , \quad m=4,5,6,7 ,
\label{sec5-20}\\
    b_\mu &=& \left\{ f\left( r, \theta \right), 0, 0, v\left( r, \theta \right) \right\} .
\label{sec5-30}
\end{eqnarray}
After substitution \eqref{sec5-10}-\eqref{sec5-30} into equations
\eqref{sec4-120}-\eqref{sec4-140} we have
\begin{eqnarray}
    \frac{1}{r^2} \frac{\partial}{\partial r}
    \left( r^2 \frac{\partial \phi}{\partial r} \right) +
    \frac{1}{r^2 \sin \theta} \frac{\partial}{\partial \theta}
    \left( \sin \theta \frac{\partial \phi}{\partial \theta}
    \right)\\
    = \phi
    \left[
      \chi^2 + \lambda_1 \left( \phi^2 - \mu^2_1 \right) -
      \left( f^2 - \frac{v^2}{r^2 \sin^2 \theta} \right)
    \right],
\label{sec5-40}\\
    \frac{1}{r^2} \frac{\partial}{\partial r}
    \left( r^2 \frac{\partial \chi}{\partial r} \right) +
    \frac{1}{r^2 \sin \theta} \frac{\partial}{\partial \theta}
    \left( \sin \theta \frac{\partial \chi}{\partial \theta} \right)
    &=& \chi
    \left[
      \phi^2 + \lambda_2 \left( \chi^2 - \mu^2_2 \right)
    \right] ,
\label{sec5-50}  \\
        \frac{1}{r^2} \frac{\partial}{\partial r}
    \left( r^2 \frac{\partial f}{\partial r} \right) +
    \frac{1}{r^2 \sin \theta} \frac{\partial}{\partial \theta}
    \left( \sin \theta \frac{\partial f}{\partial \theta} \right)
    &=& f \left(
        \frac{3}{k_2} \phi^2 - m_0^2
    \right),
\label{sec5-60}\\
    \frac{\partial^2 v}{\partial r^2} +
    \frac{\sin \theta}{r^2} \frac{\partial}{\partial \theta}
    \left( \frac{1}{\sin \theta}  \frac{\partial v}{\partial \theta} \right)
    &=& v \left(
        \frac{3}{k_2} \phi^2 - m_3^2
    \right).
\label{sec5-70}
\end{eqnarray}
The preliminary numerical investigations show that this set of
equations does not have regular solutions at arbitrary choice of
$\mu_{1,2}, m_{0,3}$ parameters. We will solve equations
\eqref{sec5-40}-\eqref{sec5-70} as a nonlinear eigenvalue problem
for eigenstates $\phi(r, \theta), \chi(r, \theta) , f(r, \theta),
v(r, \theta)$ and eigenvalues $\mu_{1,2}, m_{0,3}$. The additional
remark is that this set of equations has regular solutions not for
any values of parameters $\lambda_{1,2}$ and $k_{1,2}$. In this
paper, we take the following values: $\lambda_1 = 0.1, \lambda_2 =
1.0, k_2 = 0.5$.
\par
First, we note that the forthcoming solution depends on the
following parameters: $\phi(0), \chi(0)$. We can decrease the
number of these parameters dividing equations
\eqref{sec5-40}-\eqref{sec5-70} to $\phi^3(0)$. After this we
introduce the dimensionless radius $x=r\phi(0)$ and redefine
$\phi(r, \theta)/\phi(0) \rightarrow \phi(r, \theta), \chi(r,
\theta)/\phi(0) \rightarrow \chi(r, \theta), f(r, \theta)/\phi(0)
\rightarrow f(r, \theta), v(r, \theta)/\phi(0) \rightarrow v(r,
\theta)$ and $m_{0,3}/\phi(0) \rightarrow m_{0,3}$,
$\mu_{1,2}/\phi(0) \rightarrow \mu_{1,2}$. Thus we have the
following set of equations:
\begin{eqnarray}
    \frac{1}{x^2} \frac{\partial}{\partial x}
    \left( x^2 \frac{\partial \phi}{\partial x} \right) +
    \frac{1}{x^2 \sin \theta} \frac{\partial}{\partial \theta}
    \left( \sin \theta \frac{\partial \phi}{\partial \theta}
    \right)\\
    = \phi
    \left[
      \chi^2 + \lambda_1 \left( \phi^2 - \mu^2_1 \right) -
      \left( f^2 - \frac{v^2}{x^2 \sin^2 \theta} \right)
    \right],
\label{sec5-80}\\
    \frac{1}{x^2} \frac{\partial}{\partial x}
    \left( x^2 \frac{\partial \chi}{\partial x} \right) +
    \frac{1}{x^2 \sin \theta} \frac{\partial}{\partial \theta}
    \left( \sin \theta \frac{\partial \chi}{\partial \theta} \right)
    &=& \chi
    \left[
      \phi^2 + \lambda_2 \left( \chi^2 - \mu^2_2 \right)
    \right] ,
\label{sec5-90}  \\
        \frac{1}{x^2} \frac{\partial}{\partial x}
    \left( x^2 \frac{\partial f}{\partial x} \right) +
    \frac{1}{x^2 \sin \theta} \frac{\partial}{\partial \theta}
    \left( \sin \theta \frac{\partial f}{\partial \theta} \right)
    &=& f \left(
        \frac{3}{k_2} \phi^2 - m_0^2
    \right),
\label{sec5-100}\\
    \frac{\partial^2 v}{\partial x^2} +
    \frac{\sin \theta}{x^2} \frac{\partial}{\partial \theta}
    \left( \frac{1}{\sin \theta}  \frac{\partial v}{\partial \theta} \right)
    &=& v \left(
        \frac{3}{k_2} \phi^2 - m_3^2
    \right).
\label{sec5-110}
\end{eqnarray}
The solution of this set of equations will be regular only if
$\left.E_\theta(r, \theta)\right|_{\theta=0,\pi}=0$,
$\left.H_\theta(r, \theta)\right|_{\theta=0,\pi}=0$. The boundary
conditions will be defined below.
\par
This partial differential set of equations is extremely difficult
to solve since non-linearity and especially because of that it is
an eigenvalue problem. In order to avoid this problem we will
solve these equations approximately. In Ref. \cite{dzhun1} it is
shown that the bag formed by two equations \eqref{sec5-80}
\eqref{sec5-90} without $f, v$ is spherically symmetric. In our
situation (with four equations \eqref{sec5-80}-\eqref{sec5-110})
we will suppose that the perturbation made by the $A^8_\mu$
electromagnetic field is small enough and in the first
approximation it can be neglected. It means that in this
approximation the bag (which is described by equations
\eqref{sec5-80} \eqref{sec5-90}) remains spherically symmetric one
and only two equations \eqref{sec5-100}-\eqref{sec5-110} are
axially symmetric. Thus we have to average the term $\left( f^2 -
\frac{v^2}{r^2 \sin^2 \theta} \right)$ in the equation
\eqref{sec5-80} with respect to the angle $\theta$,
\begin{equation}
    \frac{1}{\pi} \int^{\pi}_{0} \sin\theta
    \left[ f^2\left( r,\theta \right) -
    \frac{v^2\left( r,\theta \right)}{r^2 \sin^2 \theta} \right] d\theta .
\label{sec5-120}
\end{equation}
Now we can separate the variables $r$ and $\theta$ in Eqs.
\eqref{sec5-100} \eqref{sec5-110},
\begin{eqnarray}
    f\left( r,\theta \right) &=& f(r) \Theta_1(\theta) ,
\label{sec5-130}\\
    v\left( r,\theta \right) &=& v(r) \Theta_2(\theta).
\label{sec5-140}
\end{eqnarray}
After substitution in equations \eqref{sec5-100} \eqref{sec5-110}
we obtain the following equations:
\begin{eqnarray}
    \frac{d^2 f}{dx^2} + \frac{2}{x} \frac{df}{dx} -
    \left( \frac{3}{k_2} \phi^2 - m_0^2 \right) f &=& \Lambda_1 \frac{f}{x^2} ,
\label{sec5-150}\\
    \frac{1}{\sin \theta}\frac{d}{d \theta}
    \left(
        \sin \theta \frac{d \Theta_1}{d \theta}
    \right) &=& - \Lambda_1 \Theta_1,
\label{sec5-160}\\
    \frac{d^2 v}{dx^2} -
    \left( \frac{3}{k_2} \phi^2 - m_3^2 \right) f &=& \Lambda_2 \frac{v}{x^2} ,
\label{sec5-170}\\
    \sin \theta \frac{d}{d \theta}
    \left(
        \frac{1}{\sin \theta} \frac{d \Theta_2}{d \theta}
    \right) &=& - \Lambda_2 \Theta_2 .
\label{sec5-180}
\end{eqnarray}
We take the following eigenvalues $\Lambda_{1,2}$ and eigenfunctions
$\Theta_{1,2}$
\begin{eqnarray}
    \Lambda_1 &=& 12 , \quad
    \Theta_1 = \cos \theta - \frac{5}{3} \cos^3\theta ,
\label{sec5-200}\\
    \Lambda_2  &=& 6 , \quad
    \Theta_2 = \sin^2 \theta \cos\theta
\label{sec5-220}
\end{eqnarray}
since only for this choice we will have
\begin{eqnarray}
    \left.E_\theta(r, \theta)\right|_{\theta=0,\pi} &=& 0 , \quad
    \left.H_\theta(r, \theta)\right|_{\theta=0,\pi} = 0 ,
\label{sec5-230}\\
    \left.E_r(r, \theta)\right|_{r=0} &=& 0, \quad
    \left.H_r(r, \theta)\right|_{r=0} = 0,
\label{sec5-235}\\
    M_z &\neq& 0
\label{sec5-240}
\end{eqnarray}
where $M_z$ is the total field angular momentum for the $A^8$
electromagnetic field. This choice of $\Theta_{1,2}(\theta)$ allow
us to average the equation \eqref{sec5-120},
\begin{equation}
        \frac{1}{\pi} \int^{\pi}_{0} \sin\theta
    \left[ f^2\left( x,\theta \right) -
    \frac{v^2\left( x,\theta \right)}{x^2 \sin^2 \theta} \right] d\theta =
    \frac{8}{63}f^2(x) - \frac{4}{15}\frac{v^2(x)}{x^2}.
\label{sec5-250}
\end{equation}
Finally, we have to solve the following set of equations:
\begin{eqnarray}
    \phi '' + \frac{2}{x} \phi ' &=&
    \phi \left[
        \chi^2 + \lambda_1 \left( \phi^2 - \mu_1^2 \right) -
        \left( \frac{8}{63} f^2 - \frac{4}{15} \frac{v^2}{x^2} \right)
    \right]     ,
\label{sec5-260}\\
    \chi '' + \frac{2}{x} \chi ' &=&
    \chi \left[
        \phi^2 + \lambda_2 \left( \chi^2 - \mu_2^2 \right)
    \right]     ,
\label{sec5-270}\\
    f '' + \frac{2}{x} f ' - \frac{\Lambda_1}{x^2} f &=&
    f \left( \frac{3}{k_2} \phi^2 - m_0^2 \right) ,
\label{sec5-280}\\
    v '' - \frac{\Lambda_2}{x^2} v &=&
    v \left( \frac{3}{k_2} \phi^2 - m_3^2 \right) .
\label{sec5-290}
\end{eqnarray}
The series expansions near $x=0$
\begin{eqnarray}
    \phi(x) &=& \phi_0 + \phi_3 \frac{x^2}{2} + \ldots ,
\label{sec5-292a}\\
    \chi(x) &=& \chi_0 + \chi_3 \frac{x^2}{2} + \ldots ,
\label{sec5-295a}\\
    f(x) &=& f_3 \frac{x^3}{6} + \ldots ,
\label{sec5-292}\\
    v(x) &=& v_3 \frac{x^3}{6} + \ldots
\label{sec5-295}
\end{eqnarray}
provide the constraints \eqref{sec5-230} \eqref{sec5-235}. We will
search for a regular solution with the following boundary
conditions:
\begin{eqnarray}
    \phi(0) &=& 1, \quad \phi(\infty) = \mu_1 ,
\label{sec5-296}\\
    \chi(0) &=& \chi_0, \quad \chi (\infty) = 0 ,
\label{sec5-297}\\
    f(0) &=& f(\infty) = 0,
\label{sec5-298}\\
    v(0) &=& v(\infty) = 0.
\label{sec5-299}
\end{eqnarray}
Densities of the field angular momentum and its $z-$projection are
\begin{eqnarray}
    \vec{M} &=& \left[ \vec{r} \times \left[ \vec{E} \times \vec{H} \right]\right] ,
\label{sec5-300}\\
    c M_z &=& r \sin \theta
    \left(
        H_\theta E_r - H_r E_\theta
    \right) =
\nonumber \\
    &&
    f' v' \sin^2 \theta \cos^2 \theta \left( 1 - \frac{5}{3} \cos^2 \theta \right) \\&& -
    \frac{fv}{r^2} \sin^2 \theta
    \left(
        -8 \cos^2 \theta + 15 \cos^4 \theta + 1
    \right).
\label{sec5-310}
\end{eqnarray}
The total field angular momentum is equal to
\begin{eqnarray}
    \mathcal{M}_z &=& \frac{2 \pi}{cg^2} \int_{0}^{\infty} \int_{0}^{\pi}
    r^2 \sin \theta M_z dr d \theta \nonumber \\=
    \frac{16 \pi}{105} \frac{1}{cg^2}
    \int_{0}^{\infty} x^2 \left( f' v' - 12 \frac{fv}{x^2} \right) dx ,
\label{sec5-320}\\
    \mathcal{M}_\rho &=& \mathcal{M}_\phi = 0 .
\label{sec5-315}
\end{eqnarray}
The energy density is equal to
\begin{equation}
\begin{split}
    2 \varepsilon = &
    E_i^2 + H^2_i + \left( \partial_t \phi^a \right)^2 +    \left( \partial_i \phi^a \right)^2 +
    \left( \partial_t \phi^m \right)^2 + \left( \partial_i \phi^m \right)^2 + \\
    &
    \frac{\lambda_1}{2} \left( \phi^a \phi^a -\mu_1^2 \right)^2 +
    \frac{\lambda_2}{2} \phi^m \phi^m \left( \phi^m \phi^m -2 \mu_2^2 \right) +
    \frac{k_2}{3} \left( \phi^a \phi^a  \right) \left( \phi^m \phi^m \right) - \\
    &
    \left( b_\mu b^\mu \right) \phi^a \phi^a +
    \left( m^2 \right)^{\mu \nu} b_\mu b_\nu
\label{sec5-321}
\end{split}
\end{equation}
This expression is given without the redefinition $\phi^{a,m}
\rightarrow \sqrt{\frac{3}{k_2}}\phi^{a,m}, \mu_{1,2} \rightarrow
\sqrt{\frac{3}{k_2}} \mu_{1,2}, \lambda_{1,2} \rightarrow
\frac{k_2}{3} \lambda_{1,2}$. After making this redefinition,
inserting ansatz \eqref{sec5-10}-\eqref{sec5-30} and integrating
over the angle $\theta$ yields
\begin{equation}
\begin{split}
    2g^2 \bar{\varepsilon} = & 8\pi g^2 \int^{\pi}_{0} \varepsilon \sin \theta d\theta =
    \frac{8}{63} \left( {f'}^2 + \Lambda_1 \frac{f^2}{r^2} +m_0^2 f^2 \right)\\ & +
    \frac{4}{15} \left( \frac{{v'}^2}{r^2} +
    \Lambda_2 \frac{v^2}{r^4} -m_3^2 \frac{v^2}{r^2} \right) -
    \frac{3}{k_2} \left( \frac{8}{63} f^2 - \frac{4}{15} \frac{v^2}{r^2} \right)\phi^2 \\ & +
    \frac{3}{k_2} \left[
        {\phi'}^2 + {\chi'}^2 + \frac{\lambda_1}{2} \left( \phi^2 - \mu_1^2 \right)^2 +
        \frac{\lambda_2}{2} \chi^2 \left( \chi^2 - 2 \mu_2^2 \right) +
        \phi^2 \chi^2
    \right].
\label{sec5-322}
\end{split}
\end{equation}
Here we add the constant term $\frac{\lambda_1}{2} \mu_1^4$ in
order to have a finite energy. This addition does not affect on
the field equations and gives us a finite energy of the solution.
It is necessary to note that such addition can be introduced in
the assumption \eqref{ass5} by the following scheme: $\left\langle
A^4 \right\rangle = \left\langle A^2 \right\rangle \left\langle
A^2 \right\rangle - \mu_1^2 \left\langle A^2 \right\rangle + M^4$
where $M$ is some constant which should be entered in such a way
that excludes the term $\frac{\lambda_1}{2} \mu_1^4$ in the
Lagrangian.

\subsection{Numerical calculations}

\begin{figure}[h]
  \begin{minipage}[t]{.45\linewidth}
    \centering
    \fbox{
        \includegraphics[height=5cm,width=5cm]{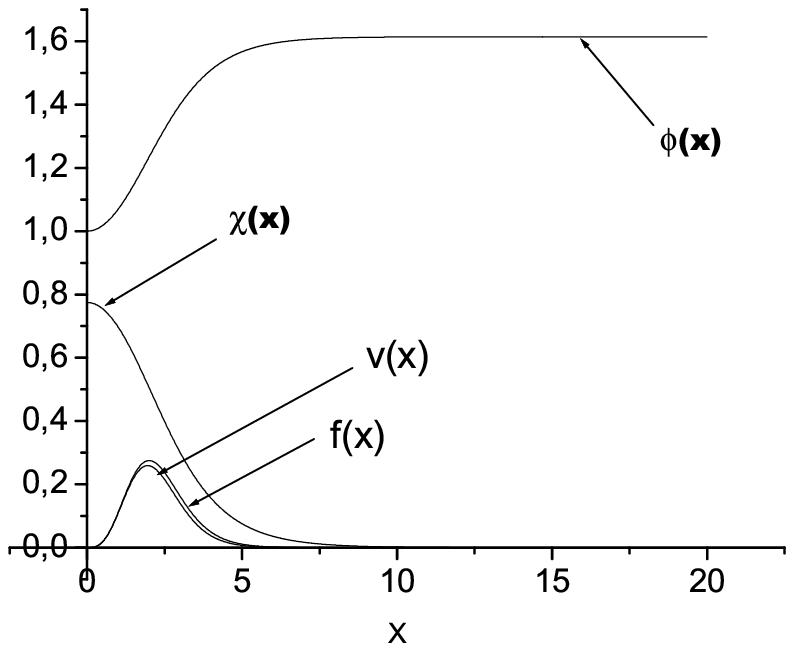}}
        \caption{The eigenfunctions $\phi(x), \chi(x), f(x), v(x)$.}
    \label{fig:functions}
  \end{minipage}\hfill
  \begin{minipage}[t]{.45\linewidth}
    \centering
    \fbox{
        \includegraphics[height=5cm,width=5cm]{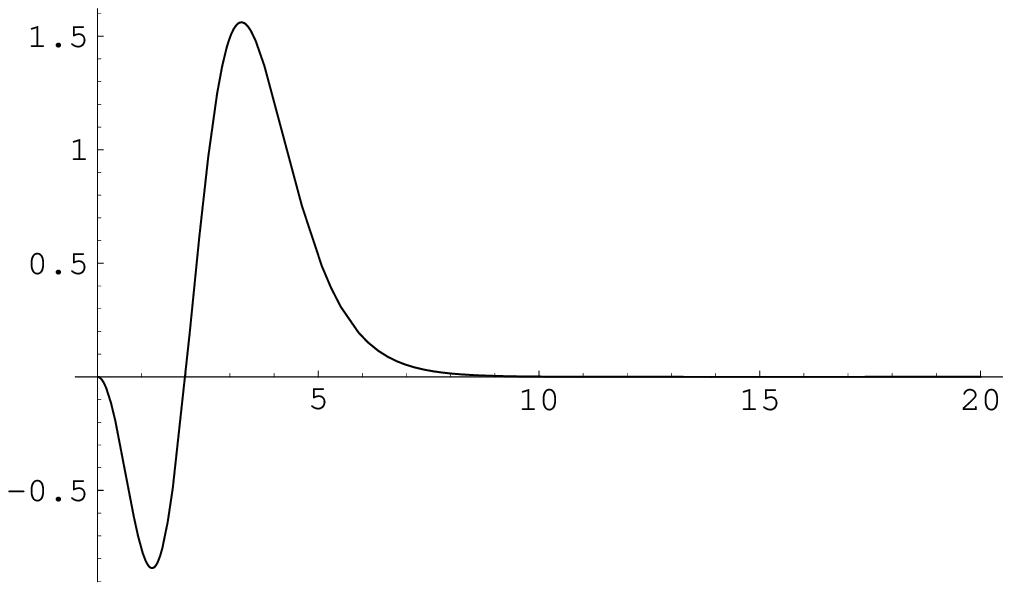}}
        \caption{The energy density $x^2 \epsilon(x)$.}
    \label{fig:energy_density}
  \end{minipage}
\end{figure}

Numerical calculations here are similar to calculations made in
Ref. \cite{dzhun1}. We search for regular solutions by shooting
method choosing $\mu_{1,2}, m_{0,3}$. The results are presented in
Fig. \eqref{fig:functions} where the eigenvalues are
\begin{equation}
    \mu_1 \approx 1.6141488, \ \mu_2 \approx 1.4925844, \
    m_0 \approx 3.6710443, \ m_3 \approx 3.46576801 .
\label{sec5-325}
\end{equation}
\par
It is easy to see that the asymptotical behaviour of the regular solution of
equations \eqref{sec5-260}-\eqref{sec5-290} is
\begin{eqnarray}
    \phi(x) \approx \mu_1 + \phi_\infty \frac{\exp\{-x\sqrt{2 \lambda_1 \mu_1^2}\}}{x},
\label{sec5-325a}\\
    \chi(x) \approx \chi_\infty \frac{\exp\{-x\sqrt{\mu_1^2 - \lambda_2 \mu_2^2}\}}{x},
\label{sec5-325b}\\
    f(x) \approx f_\infty \frac{\exp\{-x\sqrt{\frac{3}{k_2}\mu_1^2 - m_0^2}\}}{x},
\label{sec5-325c}\\
    v(x) \approx v_\infty \exp\{-x\sqrt{\frac{3}{k_2}\mu_1^2 - m_3^2}\}.
\label{sec5-325d}
\end{eqnarray}
The total field angular momentum \eqref{sec5-320} is
\begin{equation}
    \left| \mathcal{M}_z  \right|= \frac{16 \pi}{105} \frac{1}{cg^2}
    \left| I_1 \right| \approx \frac{0.46}{cg^2}
\label{sec5-330}
\end{equation}
\par
where the numerical calculations give $I_1 \approx - 0.96$. If we want to have
$\left| \mathcal{M}_z  \right| = \hbar$ then the dimensionless coupling constant
$\tilde{g}$ have to be equal to
\begin{equation}
    \tilde{g} = \frac{1/g^2}{c\hbar} \approx 2.2 .
\label{sec5-335}
\end{equation}
This quantity is equivalent to fine structure constant in quantum electrodynamic
$\alpha = e^2/(\hbar c)$ from which we immediately see that the dimensionless
coupling constant $\tilde{g} > 1$.
\par
The profile of the energy density is presented in Fig.
\eqref{fig:energy_density}. The full energy is equal to
\begin{equation}
    W = \frac{2 \pi}{g^2} \int^{\infty}_{0} \int^{\pi}_{0} r^2
    \varepsilon (r) \sin \theta dr d\theta =
    \frac{2 \pi}{g^2} \phi(0)\int^{\infty}_{0} x^2 \bar{\varepsilon} (x) dx =
    \frac{2 \pi}{g^2} \phi(0) I_2 .
\label{sec5-340}
\end{equation}
The dimensionless integral $I_2 \approx 2.75$ and consequently the full energy is
\begin{equation}
    W \approx \approx 17.3 \frac{\phi(0)}{g^2} .
\label{sec5-350}
\end{equation}
\par
The numerical analysis shows that the values of $\mu_{1,2}$ without $A^8$ field
are
\begin{equation}
    \mu_1 \approx 1.618237, \quad \mu_2 \approx 1.492871 .
\label{sec5-360}
\end{equation}
The difference between \eqref{sec5-360} and \eqref{sec5-325} is of
the order $0.2\%$. This demonstrates that $A^8$ field makes very
small perturbation of the $\left( \phi^a, \phi^m \right)$ bag and
consequently confirms our assumptions that this quantum bag
remains almost spherical one.
\par
The solution exists as well for other values of the parameters.
For example, we obtained the solution for $k_2 = 0.1$,
\begin{equation}
    \mu_1 \approx 1.6141488, \ \mu_2 \approx 1.4925844, \
    m_0 \approx 3.6710443, \ m_3 \approx 3.46576801 .
\label{sec5-370}
\end{equation}
In this case $I_1 \approx -0.13$ and
\begin{equation}
    \tilde{g} = \frac{1/g^2}{c\hbar} \approx 7.7 .
\label{sec5-380}
\end{equation}
We see that we work in the non-perturbative regime with a strong coupling
constant where the dimensionless coupling constant $\tilde{g} > 1$.
The dimensionless energy integral $I_2 \approx 21.12$ and
\begin{equation}
    W \approx 133 \frac{\phi(0)}{g^2} .
\label{sec5-390}
\end{equation}

\section{The microscopical model of inner structure of glueball with spin one}

The presented regular solution describes a quantum bag in which
the color electric and magnetic fields are confined. These fields
give an angular momentum. Thus we have a bubble of quantized and
almost-classical fields with finite energy and angular momentum
(for some choice of the parameters $\lambda_{1,2}$ and $k_2$ the
spin can be $\mathcal{M}_z = \hbar$). What is the physical
interpretation of this object? One can suppose that such an object
can be an approximate model of glueball with spin one.
\par
\begin{figure}[h]
  \begin{minipage}[t]{.45\linewidth}
    \centering
    \fbox{
        \includegraphics[height=5cm,width=5cm]{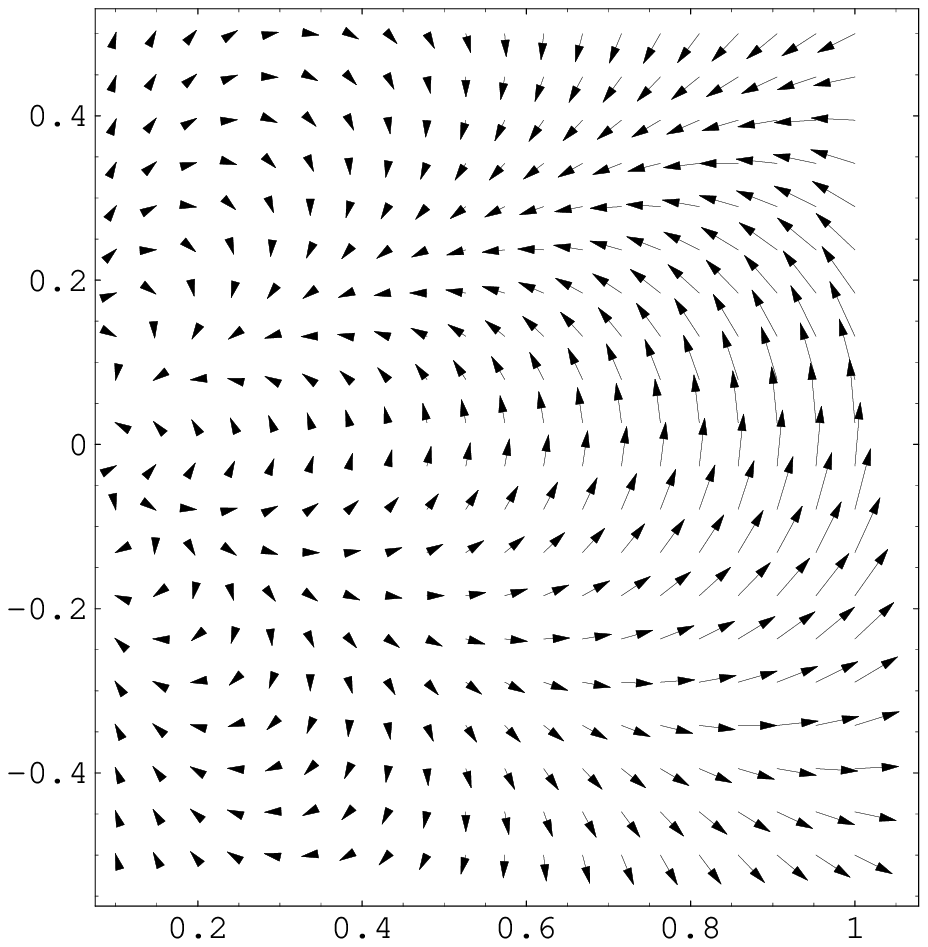}}
        \caption{The distribution of color electric field showing that near to
        the origin an electric dipole exists.}
    \label{fig:electric_field}
  \end{minipage}\hfill
  \begin{minipage}[t]{.45\linewidth}
    \centering
    \fbox{
        \includegraphics[height=5cm,width=5cm]{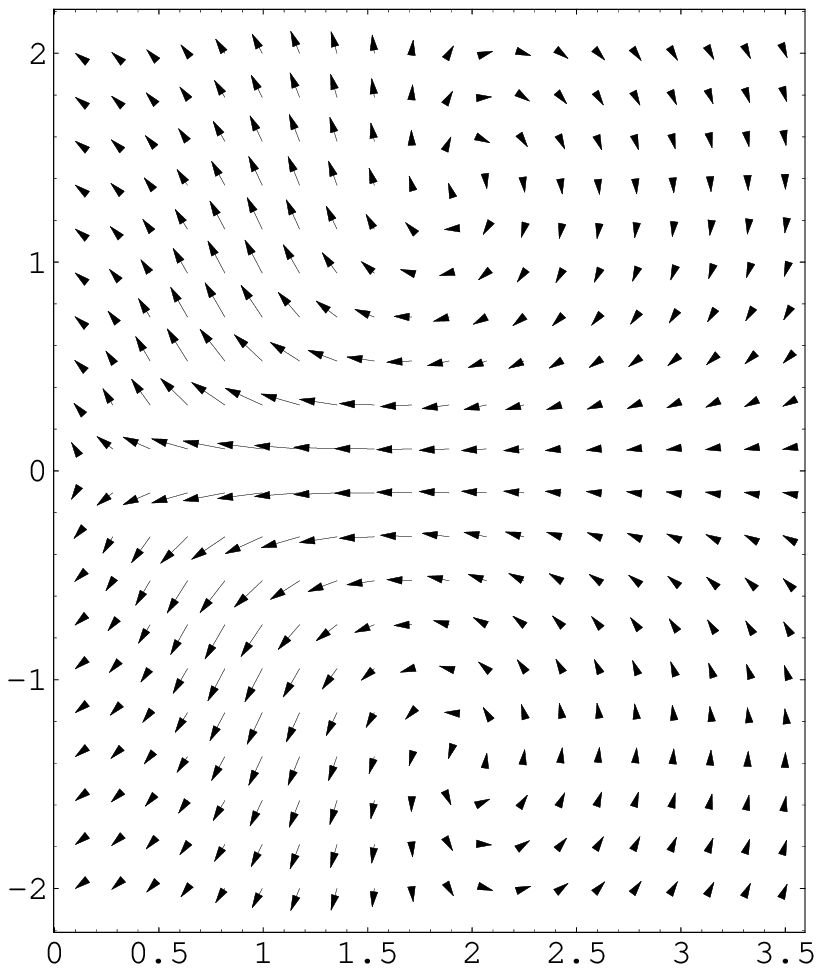}}
        \caption{The distribution of color magnetic field $A^8$ showing that two color
        electric currents exist.}
    \label{fig:magnetic_field}
  \end{minipage}
\end{figure}
\begin{figure}[h]
    \centering
    \fbox{
        \includegraphics[height=7cm,width=9cm]{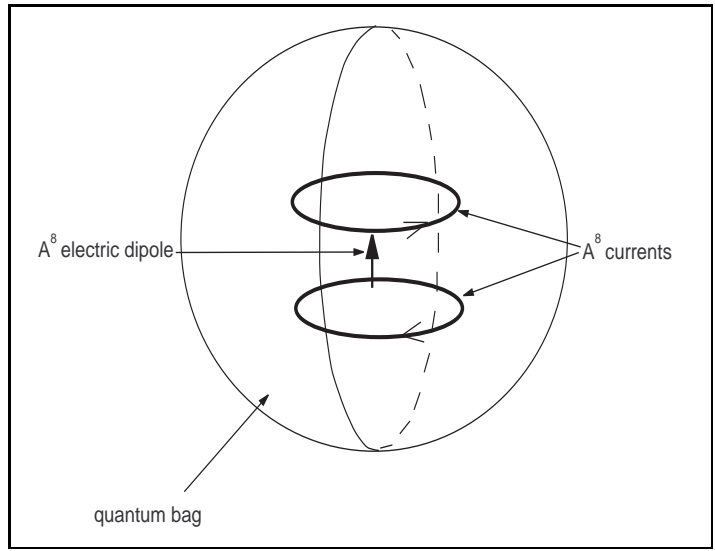}}
        \caption{The schematical picture of the bag with confined color electromagnetic
        field as the model of glueball with spin $\hbar$.}
    \label{fig:scheme_dipole_current}
\end{figure}

Now on the basis of obtained solution we would like to present the
inner structure of this object. In Fig. \eqref{fig:electric_field}
and \eqref{fig:magnetic_field} the color electric and magnetic
fields are presented.
From Fig. \eqref{fig:electric_field} one can see that at the
center there is an electric dipole and from Fig.
\eqref{fig:magnetic_field} that electric currents exist in this
object. Thus one can say that this model of glueball with spin one
approximately can be considered as the electric dipole + two
magnetic dipoles confined in a bag. Schematic view of this object
is presented in Fig. \eqref{fig:scheme_dipole_current}
\par
Let us note that similar idea was presented in Ref.
\cite{singleton}. In this notice the author shows that the
electromagnetic field angular momentum of a magnetic dipole and an
electric charge may provide a portion of the nucleon's internal
angular momentum which is not accounted for by the valence quarks.
From a rough estimate it is found that this electromagnetic field
angular momentum could contribute to the nucleon's spin $\approx
15\% \hbar$.

\section{Discussion and conclusions}

Now we would like to briefly list the results obtained above.
First, we propose a model according to which one can approximately
reduce quantum SU(3) Yang-Mills gauge theory to U(1) gauge theory
with broken gauge symmetry and interacting with scalar fields.
During such reduction the initial degrees of freedom $A^B_\mu$ are
decomposed to $A^a_\mu, A^m_\mu$ and $A^8_\mu$. $A^a_\mu$ and
$A^m_\mu$ degrees of freedom are non-perturbatively quantized in
such a way that they are similar to a non-linear oscillator, where
$\left\langle A^{a,m}_\mu \right\rangle = 0$ but $\left\langle
\left( A^{a,m}_\mu \right)^2 \right\rangle \neq 0$. $A^8_\mu$
degree of freedom remains almost classical and describe U(1) gauge
theory with broken gauge symmetry. The quantized fields $A^a_\mu,
A^m_\mu$ approximately can be described as scalar fields $\phi^m$
and $\phi^a$ correspondingly. The obtained system of field
equations has a self-consistent regular solution. Physically this
solution presents a pure quantum bag which is described by
interacting fields $\phi^m$ and $\phi^a$ and electromagnetic field
$A^8_\mu$ which is confined inside the bag. The color
electromagnetic field $A^8_\mu$ is the source of an angular
momentum. The obtained object is a cloud of quantized fields with
a spin (for some values of $\lambda_{1,2}, k_2$ the spin can be
equal to $\hbar$). We suppose that such an object can be
considered as a model of glueball with spin one. The presence of
the color electromagnetic field $A^8_\mu$ leads to an asymmetrical
structure in the glueball and nucleon that probably can be proved
experimentally.
\par
Summarizing the results of this paper and Refs. \cite{dzhun1},
\cite{dzhun2} one can say that the interaction between $A^a_\mu$
and $A^m_\mu$ degrees of freedom gives a quantum bag. If $A^a_\mu$
is non-quantized, in the result we have a flux tube with a
longitudinal color electric field. If these degrees of freedom are
quantized we have a quantum bag which can be considered as a model
of glueball with spin zero. This paper and Ref. \cite{dzhun3} show
correspondingly that this bag can sustain an electromagnetic field
and colorless spinor field (one can say that the bag is strong
enough).
\par
In this paper, we have shown that the interaction between the
condensates $\left\langle A^a_\mu A^{a\mu} \right\rangle$,
$\left\langle A^m_\mu A^{m\mu} \right\rangle$ and $\left\langle
b_\mu b^\mu \right\rangle$ is a necessary condition for the
existence of the quantum bag where the field $A^8$ is confined.
One important thing here is that these calculations are
non-perturbative and do not use Feynman diagram.

\section*{Acknowledgment}

I am very grateful to the Alexander von Humboldt Foundation for the financial support
and thanks Prof. H. Kleinert for hospitality in his research group.


\begin{thebibliography}{99}

\bibitem{dzhun1}
V. Dzhunushaliev, ``Scalar model of the glueball'', to appear in
Hadronic J., hep-ph/0312289.

\bibitem{Simonov:2005ir}
Y.~A.~Simonov, ``Analytic calculation of field-strength
correlators'', hep-ph/0501182.

\bibitem{Gripaios}
B. M. Gripaios, Phys. Lett. B \textbf{558}, 250 (2003).

\bibitem{Kondo}
K.-I. Kondo, Phys. Lett. B \textbf{572}, 210 (2003).

\bibitem{Slavnov}
A. A. Slavnov, hep-th/0407194.

\bibitem{Stodolsky}
L. Stodolsky, Pierre van Baal and V. I. Zakharov,
Phys. Lett. B \textbf{552}, 214(2002).

\bibitem{Gubarev}
F. V. Gubarev, L. Stodolsky, and V. I. Zakharov, Phys. Rev.
Lett. \textbf{86}, 2220 (2001).

\bibitem{Zakharov}
F. V. Gubarev, V. I. Zakharov, Phys. Lett. B \textbf{501}, 28
(2001).

\bibitem{Gracey}
J. A. Gracey, Phys. Lett. B
\textbf{552}, 101 (2003).

\bibitem{Verschelde}
H. Verschelde, K. Knecht, K. Van Acoleyen and V. Vanderkelen,
Phys. Lett. B \textbf{516}, 307 (2001).

\bibitem{Dudal}
D. Dudal, H. Verschelde, J. A. Gracey, V. E. R. Lemes, M. S.
Sarandy, R.~F.~Sobreiro and S. P. Sorella, JHEP \textbf{0401}, 044
(2004); hep-th/0311194 v3.

\bibitem{heis}
W. Heisenberg, \textit{Introduction to the unified field theory of
elementary particles.}, Max - Planck - Institut f\"ur Physik und
Astrophysik, Interscience Publishers London, New York, Sydney,
1966; W. Heisenberg, Nachr. Akad. Wiss. G\"ottingen, N8, 111
(1953); W. Heisenberg, Zs. Naturforsch., \textbf{9a}, 292 (1954);
W. Heisenberg, F. Kortel und H. M\"utter, Zs. Naturforsch.,
\textbf{10a}, 425 (1955); W. Heisenberg, Zs. f\"ur Phys.,
\textbf{144}, 1 (1956); P. Askali and W. Heisenberg, Zs.
Naturforsch., \textbf{12a}, 177 (1957); W. Heisenberg, Nucl.
Phys., \textbf{4}, 532 (1957); W. Heisenberg, Rev. Mod. Phys.,
\textbf{29}, 269 (1957).

\bibitem{giacomo}
A. Di Giacomo, H.G. Dosch, V.I. Shevchenko and Yu. A. Simonov,
%``Field correlators in QCD. Theory and applications'',
Phys. Rep., \textbf{372}, 319 (2002), hep-ph/0007223.

\bibitem{simonov}
Yu. A. Simonov,
``Selfcoupled equations for the field correlators'',
hep-ph/9712250.

\bibitem{Li:2004zu}
X.~d.~Li and C.~M.~Shakin,
%``Quark propagation in the presence of a  condensate,''
Phys. Rev. D {\bf 70} (2004) 114011.

\bibitem{singleton}
D. Singleton, Phys. Lett. \textbf{B427}, 155 (1998).

\bibitem{dzhun2}
V. Dzhunushaliev, ``The colored flux tube'' to appear in Hadronic
J., hep-ph/0307274.

\bibitem{dzhun3}
V. Dzhunushaliev, ``Glueball filled with quark field as a model of
nucleon'' to appear in Hadronic J., hep-ph/0408236.


\end{thebibliography}
\end{document}